\documentclass[preprint, superscriptaddress,showpacs,preprintnumbers,amsmath,amssymb]{revtex4}
\usepackage{graphicx}

\begin{document}

\thispagestyle{empty}

\title{Casimir free energy and pressure for magnetic metal films}

\author{
G.~L.~Klimchitskaya}

\author{
V.~M.~Mostepanenko}
\affiliation{Central Astronomical Observatory at Pulkovo of the Russian Academy of Sciences,
Saint Petersburg,
196140, Russia}
\affiliation{Institute of Physics, Nanotechnology and
Telecommunications, Peter the Great Saint Petersburg
Polytechnic University, St.Petersburg, 195251, Russia}

\begin{abstract}
We examine the Casimir free energy and pressure of magnetic metal films, which are
free-standing in vacuum, sandwiched between two dielectric plates, and deposited on
either nonmagnetic or magnetic metallic plates. All calculations are performed using
both the Drude and plasma model approaches to the Lifshitz theory. According to our
results, the Casimir free energies and pressures calculated using both theoretical
approaches are significantly different in the magnitude and sign even for thin films
of several tens of nanometers thickness. Thus, for the Ni film of 47\,nm thickness
deposited on a Fe plate the obtained magnitudes of the Casimir free energy differ by
the factor of 5866. We show that the Casimir free energy and pressure of a magnetic
film calculated using the plasma model approach do not possess the classical limit,
but exponentially fast drop to zero with increasing film thickness. If the Drude
model approach is used, the classical limit is reached for magnetic films of about
150\,nm thickness, but the Casimir free energy remains nonzero in the limit of ideal
metal, contrary to expectations. For the plasma model approach the Casimir free energy
of a film vanishes in this case. Numerical computations are performed for the magnetic
films made of Ni, nonmagnetic plates made of Cu and Al, and magnetic plates made of Fe
using the tabulated optical data for the complex indexes of refraction of all metals.
The obtained results can be used for a discrimination between the plasma and Drude
model approaches in the Casimir physics and in the investigation of stability of thin
films.
\end{abstract}

\maketitle

\section{Introduction}

The physical phenomena caused by the electromagnetic fluctuations attract increasing attention because
they are closely connected with the basics of quantum statistical physics and simultaneously are
important for many innovative technologies. The most well known fluctuation-induced phenomenon is
the van der Waals force \cite{1} acting at short separations between two uncharged particles,
a particle and a macrobody and between two macrobodies. Its relativistic analogue, the Casimir
force \cite{2}, has also received much publicity in connection with the zero-point fluctuations
of quantized fields, although this effect originates from the thermal fluctuations as well.
Both forces are often mentioned under a generic name of dispersion forces.

In spite of many applications of the Casimir effect in atomic physics \cite{3,4,5,6,7,8},
condensed matter physics \cite{9,10,11,12,13,14} and nanotechnology \cite{15,16,17,18,19},
the most important outcome is, probably, that it sheds new light on the interaction of
quantum fluctuations with matter and the role of dissipation in this process \cite{20}.
Specifically, it was shown \cite{21,22,23,23a,24} that if the dielectric response of metals with
perfect crystal lattices is described by the dissipative Drude model (as is certainly true in
real electromagnetic fields with a nonzero expectation value, but is usually applied to fluctuating
fields with zero expectation value as well)
the Lifshitz theory of dispersion forces comes into conflict with the third
law of thermodynamics, the Nernst heat theorem.
Note that when the spatially nonlocal dielectric response is
considered, the Nernst heat theorem is satisfied \cite{25a}.
The reason is that effects of spatial dispersion lead to an
effective nonzero residual relaxation. This, however, does not
solve the problem. The point is that at sufficiently short
separations between the Casimir plates the frequency region of
infrared optics, where the dielectric response is local, plays an
important role. It was shown that if the frequency regions with
both nonlocal and local response functions are taken into account,
 the Nernst heat theorem is again violated \cite{25b}.

In parallel with this, several precise
experiments performed by two experimental groups demonstrated that measurements of the Casimir
interaction between both nonmagnetic and magnetic test bodies exclude theoretical predictions
of the Lifshitz theory taking into account dissipation of free electrons at up to 99\%
confidence level \cite{25,26,27,28,29,30,31}. The same experimental data were found in
agreement with the Lifshitz theory if the free electrons are described by the nondissipative
plasma model \cite{25,26,27,28,29,30,31} at more than  90\% confidence level \cite{32}.
Note that the magnetic response falls so rapidly with increasing frequencies \cite{53}
that the magnetic permeability is effectively unity for all nonzero-frequency Matsubara terms
of the Lifshitz formula and takes its static value for the zero-frequency Matsubara term
only \cite{54}.
It was shown also \cite{21,22,23,23a,24} that the Lifshitz theory satisfies the Nernst theorem
for both nonmagnetic and magnetic materials
if the plasma model is used in place of the Drude one.

At separations below $1\,\mu$m between the test bodies, where the experiments of
Refs.~\cite{25,26,27,28,29,30,31} are the most precise, the differences in theoretical predictions
using the Drude and plasma models do not exceed a few percent. Because of this, it has been
argued on occasion that the obtained results might be not conclusive. Recently, however, several
differential Casimir experiments exploiting magnetic metals have been proposed \cite{33,34,35},
where the theoretical predictions
using the Drude and plasma models at low frequencies differ by up to a factor of 1000.
One of these experiments \cite{36,36a} has already demonstrated with certainty that the Drude
model approach is excluded by the data, whereas the plasma model approach was found to be
consistent with the same data.

In addition to experiments with metallic test bodies, several precision measurements of the
Casimir and Casimir-Polder force with dielectric test bodies have been made  \cite{37,38,39,40}.
The experimental data were found in agreement with theoretical predictions of the Lifshitz
theory only under a condition that the contribution of free charge carriers to the
dielectric permittivity is disregarded in calculations. If the free charge carriers are taken
into account, the computational results are excluded by the data at up to 95\%
confidence level \cite{37,38,39,40,41}. It was proved also that for dielectric materials the
Lifshitz theory violates the Nernst heat theorem if the free charge carriers at nonzero
temperature are taken into account and satisfies it if the free charge carriers are disregarded
\cite{23a,42,43,44,45,46}.

Recently a new type of the Casimir experiments was proposed which allows
an immediate discrimination
between the Drude and plasma model approaches without resorting to the differential force
measurements \cite{47}. It was shown \cite{47} that the Casimir free energy and pressure of
a nonmagnetic metallic film with less than 100\,nm thickness can differ up to a factor of
several tens when it is calculated using different theoretical approaches. In so doing, this film
can be either a free-standing or sandwiched between two dielectric plates.
Similar results have been obtained for nonmagnetic metal films deposited on the plates
made of nonmagnetic metals \cite{49a}.
Note that the clarification
of a question what is the van der Waals and Casimir energy of metallic films is interesting not
only from the theoretical point of view. The matter is that these interactions are important for
the problem of stability of thin films and should be taken into account in numerous
applications \cite{48}.

In this paper, we investigate the Casimir free energy and pressure of thin films made of magnetic
metal. Interest to this subject is motivated by the recent experiments on measuring the Casimir force
between magnetic test bodies \cite{30,31,36,36a,49} and the great role played by magnetic
coatings in various applications. We consider the Casimir free energy and pressure of a magnetic
film either a free-standing or sandwiched between two dielectric plates. We also calculate
the Casimir free energy and pressure of magnetic films deposited on either nonmagnetic or
magnetic metal plates. All calculations are performed using both the Drude and plasma model
approaches, and in all cases important differences in the obtained results are revealed.

Specifically, if the Drude model approach is used, the Casimir free energy and pressure have
the classical limit, which is already reached for thin films of about 150\,nm thickness.
Recall that in the classical limit the Casimir free energy and pressure become essentially
classical, i.e., the main contributions to them do not depend on the Planck constant.
For two metallic plates separated by a vacuum gap the classical limit is reached at
much larger separations (higher temperatures).
The Casimir free energy and pressure of a film calculated using the plasma model approach
do not have the classical limit, i.e., the main contributions to them remain dependent on
the Planck constant for any film thickness. In this case the Casimir free energy and pressure
 decrease to zero exponentially fast with increasing thickness of the film.
Because of this, even for thin films of a few tens of nanometers thickness,
the difference in theoretical predictions of both approaches reaches hundreds and thousands
percent. This allows an experimental discrimination between different predictions do not
using the differential force measurements. According to our results, in the limiting case
of ideal metal the Casimir free energy and pressure of a magnetic metal film calculated using
the Drude model approach do not go to zero. It is in conflict with intuition if to take into
account that the electromagnetic fluctuations cannot penetrate in an interior of ideal
metal. The plasma model approach is shown to be in accordance with this demand.

Numerical computations are performed for a Ni film free-standing in vacuum or sandwiched
between two sapphire plates. We have also computed the Casimir free energy and pressure of
the magnetic Ni film deposited on a nonmagnetic plate made of Cu or Al and a magnetic plate
made of Fe. All computations are performed using the complete optical data of all metals under
consideration extrapolated to zero frequency by either the Drude or the plasma model. It is shown that
for Ni films deposited on Al or Cu plates the Casimir free energy and pressure calculated using
the plasma model approach are positive (i.e., correspond to the Casimir repulsion). If the Drude model
approach is used, the Casimir free energy and pressure change their sign with increasing film thickness.
For a Ni film deposited on a Fe plate the Casimir free energy and pressure are positive if
the Drude model approach is used in calculations and change their sign with increasing film thickness
if one uses the plasma model approach.

The paper is organized as follows. In Sec.~II the general formalism is presented and the
dielectric functions of all considered metals along the imaginary frequency axis are displayed.
Section~III is devoted to a magnetic metal film in vacuum or sandwiched between two dielectrics.
In Sec.~IV we consider a magnetic metal film deposited on nonmagnetic metal plates.
In Sec.~V a magnetic metal film deposited on a magnetic metal plate is considered.
Section~VI contains our conclusions and discussion.

\section{General formalism}

We consider a magnetic metal film of thickness $a$ characterized by the frequency-dependent
dielectric permittivity $\varepsilon^{(2)}(\omega)$ and magnetic permeability $\mu^{(2)}(\omega)$.
This film may be sandwiched between two thick plates (semispaces) described by the
quantities $\varepsilon^{(1)}(\omega),\,\,\mu^{(1)}(\omega)$ and
$\varepsilon^{(3)}(\omega),\,\,\mu^{(3)}(\omega)$.
If these plates are nonmagnetic, the equalities $\mu^{(1)}(\omega)=\mu^{(3)}(\omega)=1$ hold.
For a free-standing film in vacuum one also has
 $\varepsilon^{(1)}(\omega)=\varepsilon^{(3)}(\omega)=1$.
 Finally, for a film deposited on a plate we have
$\varepsilon^{(3)}(\omega)=\mu^{(3)}(\omega)=1$.
The Casimir free energy per unit area of our magnetic film at temperature $T$ in thermal
equilibrium with an environment is given by the Lifshitz formula \cite{1,2,51,52}
\begin{equation}
{\cal F}(a,T)=\frac{k_BT}{2\pi}\sum_{l=0}^{\infty}{\vphantom{\sum}}^{\prime}
\int_{0}^{\infty}k_{\bot}\,dk_{\bot}
\sum_{\alpha}\ln\left[1-r_{\alpha}^{(2,3)}(i\xi_l,k_{\bot})
r_{\alpha}^{(2,1)}(i\xi_l,k_{\bot})e^{-2ak_l^{(2)}(k_{\bot})}\right],
\label{eq1}
\end{equation}
\noindent
where, $k_B$ is the Boltzmann constant,
$k_{\bot}=|\mbox{\boldmath$k$}_{\bot}|$ is the magnitude of the
projection of the wave vector on the surface of the film, and
the prime on a summation sign over $l$ multiplies the term with $l=0$
by 1/2. The second summation is over two independent
polarizations of the electromagnetic field, transverse magnetic
($\alpha={\rm TM}$) and transverse electric ($\alpha={\rm TE}$).
The respective reflection coefficients calculated at the pure imaginary
Matsubara frequencies
$\xi_l=2\pi k_BTl/\hbar$, $l=0,\,1,\,2,\,\ldots$,
are given by
\begin{eqnarray}
&&
r_{\rm TM}^{(2,n)}(i\xi_l,k_{\bot})=\frac{\varepsilon_{l}^{(n)}
k_{l}^{(2)}(k_{\bot})-\varepsilon_{l}^{(2)}
k_l^{(n)}(k_{\bot})}{\varepsilon_{l}^{(n)}
k_{l}^{(2)}(k_{\bot})+\varepsilon_{l}^{(2)}
k_l^{(n)}(k_{\bot})},
\nonumber \\
&&
r_{\rm TE}^{(2,n)}(i\xi_l,k_{\bot})=\frac{\mu_l^{(n)}k_{l}^{(2)}(k_{\bot})-
\mu_l^{(2)}k_l^{(n)}(k_{\bot})}{\mu_l^{(n)}k_{l}^{(2)}(k_{\bot})+
\mu_l^{(2)}k_l^{(n)}(k_{\bot})},
\label{eq2}
\end{eqnarray}
\noindent
where
\begin{equation}
k_l^{(n)}(k_{\bot})=\sqrt{k_{\bot}^2+
\mu_l^{(n)}{\varepsilon_{l}^{(n)}}\frac{\xi_l^2}{c^2}},
\label{eq3}
\end{equation}
\noindent
 $\mu_{l}^{(n)}\equiv\mu^{(n)}(i\xi_l)$,
  $\varepsilon_{l}^{(n)}\equiv\varepsilon^{(n)}(i\xi_l)$,
and $n =1,\,2,\,3$.

As is mentioned in Sec.~I,
the magnetic permeabilities of metals along the imaginary frequency axis
decrease with increasing frequency and at room temperature  $T=300\,$K drop
to unity at much lower frequencies than the first Matsubara frequency \cite{53,54}.
This means that in all terms of Eq.~(\ref{eq1}) with $l\geq 1$ one can put
$\mu_{l}^{(n)}=1$. As a result, in all terms of Eq.~(\ref{eq1}) with $l\geq 1$
the reflection coefficients preserve their form (\ref{eq2}) where one should put
\begin{equation}
k_l^{(n)}(k_{\bot})=\sqrt{k_{\bot}^2+
{\varepsilon_{l}^{(n)}}\frac{\xi_l^2}{c^2}}, \quad
\mu_l^{(n)}=1.
\label{eq4}
\end{equation}
\noindent
The reflection coefficients at $l=0$ take different forms depending on the
configuration under consideration. They are presented in the following sections.
Here we only notice that for the magnetic films made of Ni and magnetic plates made
of Fe we have \cite{55}
\begin{equation}
\mu_0^{(2)}\equiv\mu_{\rm Ni}=110, \quad
\mu_0^{(1)}\equiv\mu_{\rm Fe}=10^4.
\label{eq5}
\end{equation}

To perform computations of the Casimir free energy using Eq.~(\ref{eq1}), one also
needs the values of dielectric permittivities at the imaginary Matsubara frequencies.
In the next sections we consider Ni and Pt films and Al, Cu, and Fe plates.
The dielectric permittivities for all of them along the imaginary frequency axis were
obtained from the tabulated optical data for the complex index of refraction \cite{50}
using the Kramers-Kronig relation. The respective procedure is described in detail
in Refs.~\cite{2,52} for the case of Au.
In the literature, there are some alternative sets of optical data for metal films
which also depend on a specific sample used. Our results below, however, are almost
independent on what set of optical data is used in computations.
 An extrapolation of the optical data down to
zero frequency was performed using both the Drude and plasma models
\begin{equation}
\varepsilon_D(i\xi)=1+
\frac{\omega_{p}^2}{\xi_l(\xi_l+\gamma)},\quad
\varepsilon_p(i\xi)=1+
\frac{\omega_{p}^2}{\xi_l^2},
\label{eq6}
\end{equation}
\noindent
where $\omega_{p}$ is the plasma frequency and $\gamma$ is the relaxation
parameter, for the reasons described in Sec.~I.
Note that both these dielectric permittivities considered in the complex frequency plane
satisfy the Kramers-Kronig relations formulated for functions having the first- and
second-order poles at zero frequency, respectively \cite{2,52}.

\begin{figure}[b]
\vspace*{-2cm}
\centerline{\hspace*{0cm}
\includegraphics{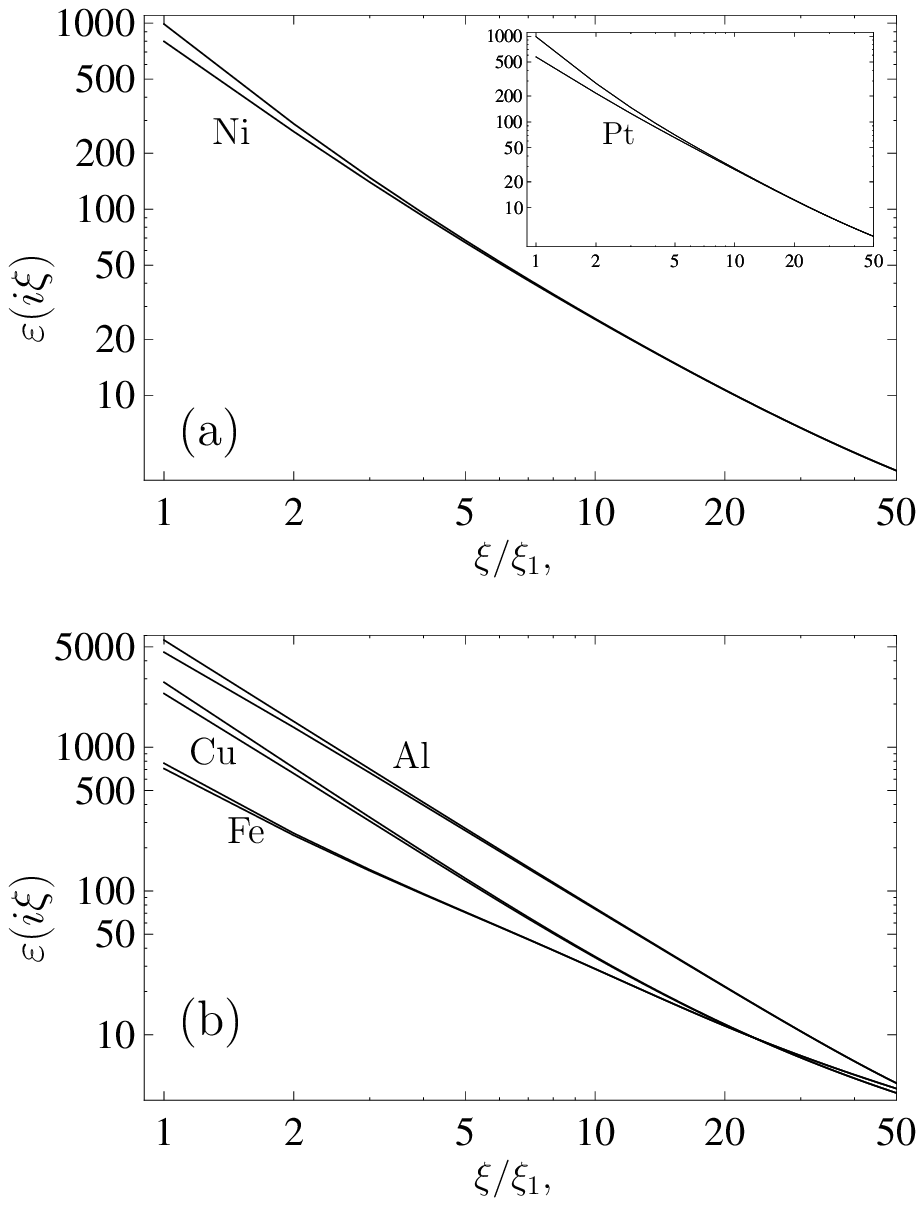}
}
\vspace*{-13cm}
\caption{\label{fg1}
The dielectric permittivities of (a) Ni and Pt (the inset) and (b) Al, Cu, and Fe
are shown as functions of the imaginary frequency normalized to the first
Matsubara frequency by the pairs of upper and lower lines obtained through
extrapolations of the optical data
using the plasma and Drude models, respectively.
}
\end{figure}

In Fig.~\ref{fg1}(a) the obtained dielectric permittivities of Ni and Pt (in the inset) are
shown as functions of the imaginary frequency normalized to the first Matsubara frequency
at room temperature
$\xi_1=2.468\times 10^{14}\,\mbox{rad/s}=0.1624\,$eV.
The upper and lower solid lines plotted in the double logarithmic scale are obtained using
the plasma and Drude extrapolations, respectively. The plasma frequencies and relaxation
parameters used in the extrapolations are
$\omega_{p,{\rm Ni}}=4.89\,$eV, $\gamma_{\rm Ni}=0.0436\,$eV \cite{50,56}, and
$\omega_{p,{\rm Pt}}=4.94\,$eV, $\gamma_{\rm Pt}=0.13\,$eV (close to Ref.~\cite{56}),
respectively. For Al, Cu, and Fe the dielectric permittivities along the imaginary frequency
axis are presented in Fig.~\ref{fg1}(b) in a similar way. The following
plasma frequencies and relaxation parameters were used in the extrapolations:
$\omega_{p,{\rm Al}}=11.34\,$eV, $\gamma_{\rm Al}=0.041\,$eV (close to Ref.~\cite{57}),
$\omega_{p,{\rm Cu}}=8.6\,$eV, $\gamma_{\rm Cu}=0.0325\,$eV (close to Ref.~\cite{56}), and
$\omega_{p,{\rm Fe}}=4.09\,$eV, $\gamma_{\rm Fe}=0.018\,$eV \cite{56}. As is seen
in Fig.~\ref{fg1}(a,b), the most important impact from different extrapolations is given
only by the first Matsubara frequency.

In the next section we also consider the magnetic Ni film sandwiched between two sapphire
plates. The dielectric permittivity of sapphire along the imaginary frequency
axis admits rather precise analytic expression \cite{58}
\begin{equation}
\varepsilon_s(i\xi)= 1+\frac{C_{\rm IR}\omega_{\rm IR}^2}{\omega_{\rm IR}^2+\xi^2}
+\frac{C_{\rm UV}\omega_{\rm UV}^2}{\omega_{\rm UV}^2+\xi^2},
\label{6a}
\end{equation}
\noindent
where
$C_{\rm UV}=2.072$, $C_{\rm IR}=7.03$,
$\omega_{\rm UV}=2.0\times 10^{16}\,$rad/s,
and $\omega_{\rm IR}=1.0\times 10^{14}\,$rad/s.
This permittivity was recently used in
Ref.~\cite{47} and is used below as well.

In the end of this section we present the Lifshitz formula for the Casimir pressure for
a magnetic metal film included in the same configurations, as described above.
It is a direct consequence of Eq.~(\ref{eq1}):
\begin{equation}
P(a,T)=-\frac{k_BT}{\pi}\sum_{l=0}^{\infty}{\vphantom{\sum}}^{\prime}
\int_{0}^{\infty}k_{\bot}k_l^{(2)}(k_{\bot})\,dk_{\bot}
\sum_{\alpha}\left[
\frac{e^{2ak_l^{(2)}(k_{\bot})}}{r_{\alpha}^{(2,3)}(i\xi_l,k_{\bot})
r_{\alpha}^{(2,1)}(i\xi_l,k_{\bot})}
-1\right]^{-1\!\!\!},
\label{eq7}
\end{equation}
\noindent
where the reflection coefficients are given by Eq.~(\ref{eq2}) taking into account
Eq.~(\ref{eq4}) at $l\geq 1$.

\section{Magnetic film sandwiched between two dielectrics}

Here, we consider the configuration of a magnetic (Ni) film of thickness $a$ either
sandwiched between two thick plates made of a nonmagnetic dielectric with some
static dielectric permittivity $\varepsilon_0^{(1)}=\varepsilon_0^{(3)}$ (sapphire)
or free-standing in vacuum (the latter can also be considered as a pair of dielectric
semispaces). Before starting with numerical computations, we consider the contributions
of the zero-frequency terms in Eqs.~(\ref{eq1}) and (\ref{eq7}) because this alone leads
to some important conclusions of qualitative character.

If the dielectric permittivity of Ni is described by the Drude model [the first equality in
Eq.~(\ref{eq6})], Eq.~(\ref{eq3}) leads to $k_0^{(n)}(k_{\bot})=k_{\bot}$ and Eq.~(\ref{eq2})
results in
\begin{eqnarray}
&&
r_{{\rm TM},D}^{(2,n)}(0,k_{\bot})=-1,
\label{eq8} \\
&&
r_{{\rm TE},D}^{(2,n)}(0,k_{\bot})=\frac{1-\mu_{\rm Ni}}{1+\mu_{\rm Ni}}
\equiv r_{0,{\rm Ni}}.
\nonumber
\end{eqnarray}
\noindent
As is seen from Eq.~(\ref{eq8}), the zero-frequency reflection coefficients calculated using
the Drude model do not depend on a material of dielectric plates, i.e., do not depend on
$\varepsilon_{0}^{(1)}$. In the case of a free-standing magnetic film in vacuum
$\varepsilon_{0}^{(1)}=1$.

Substituting  Eq.~(\ref{eq8}) in Eq.~(\ref{eq1}), for the contribution of zero Matsubara
frequency to the Casimir free energy calculated using the Drude model approach one finds
\begin{equation}
{\cal F}_{D}^{(l=0)}(a,T)=\frac{k_BT}{4\pi}\int_{0}^{\infty}k_{\bot}dk_{\bot}
\left[\ln(1-e^{-2ak_{\bot}})+\ln(1-r_{0,{\rm Ni}}^2e^{-2ak_{\bot}})\right].
\label{eq9}
\end{equation}
\noindent
In terms of the dimensionless integration variable $y=2ak_{\bot}$ Eq.~(\ref{eq9})
takes the form
\begin{equation}
{\cal F}_{D}^{(l=0)}(a,T)=\frac{k_BT}{16\pi a^2}\int_{0}^{\infty}ydy
\left[\ln(1-e^{-y})+\ln(1-r_{0,{\rm Ni}}^2e^{-y})\right].
\label{eq10}
\end{equation}
\noindent
Calculating both integrals on the right-hand side on this equation, we arrive at
\begin{equation}
{\cal F}_{D}^{(l=0)}(a,T)=-\frac{k_BT}{16\pi a^2}\left[\zeta(3)+
{\rm Li}_3(r_{0,{\rm Ni}}^2)\right],
\label{eq11}
\end{equation}
\noindent
where $\zeta(z)$ is the Riemann zeta function and ${\rm Li}_n(z)$ is the polylogarithm
function.

The first term on the right-hand side of Eq.~(\ref{eq11}) coincides with that obtained
earlier for a nonmagnetic metal film \cite{47}, whereas the second describes the contribution
of magnetic properties. Taking into account that $r_{0,{\rm Ni}}\approx -1$, the magnetic
properties increase the magnitude of ${\cal F}_{D}^{(l=0)}$ by approximately a factor of two.
It is well known \cite{2} that for two plates separated by a vacuum gap of width $a$ the
zero-frequency term gives the major contribution to the Casimir free energy in the
classical limit which is reached at $a>6\,\mu$m.
As is seen from numerical computations below, for Ni films described by the Drude model
approach the classical limit is already reached for relatively small film thicknesses of
about 150\,nm.
Note that Eq.~(\ref{eq11}) presents the nonzero Casimir free energy of a metal film in the
limiting case of an ideal metal \cite{47}.
We recall that for an ideal metal (or, synonymously, for a perfect reflector) the
magnitudes of both the TM and TE reflection coefficients must be equal to unity at
all frequencies. This is achieved when
 $\omega_p\to\infty$ and one obtains Eq.{\ }(\ref{eq11}).
 The obtained result is
in contradiction with the fact that the electromagnetic oscillations cannot penetrate in an
interior of ideal metal. Previously Refs.~\cite{60a,60b,60c} discussed similar problems
concerning the noncommutativity of the limiting transitions to zero frequency and
infinitely large dielectric permittivity.

In a similar way, substituting  Eq.~(\ref{eq8}) in Eq.~(\ref{eq7}), for the zero-frequency
contribution  to the Casimir pressure calculated using the Drude model approach one obtains
\begin{equation}
{P}_{D}^{(l=0)}(a,T)=-\frac{k_BT}{8\pi a^3}\left[\zeta(3)+
{\rm Li}_3(r_{0,{\rm Ni}}^2)\right].
\label{eq14}
\end{equation}
\noindent
This is also obtainable by the negative differentiation with respect to $a$ from the free
energy (\ref{eq11}), as it should be. The role of magnetic properties in
Eq.~(\ref{eq17}) is the same, as already discussed above with respect to Eq.~(\ref{eq11}).

Unlike the case of two plates separated with a gap of width $a$, an application of the
plasma model approach to a magnetic film of thickness $a$ leads to totally different
results for already small $a$ of several tens of nanometers. Although for the TM reflection
coefficient at zero Matsubara frequency Eq.~(\ref{eq2}) leads to
\begin{equation}
r_{{\rm TM},p}^{(2,n)}(0,k_{\bot})=-1,
\label{eq15}
\end{equation}
\noindent
i.e., to the same result as the Drude model, for the TM reflection
from Eq.~(\ref{eq2}) we obtain
\begin{equation}
r_{{\rm TE},p}^{(2,n)}(0,k_{\bot})=
\frac{\sqrt{c^2k_{\bot}^2+\mu_{\rm Ni}\omega_{p,{\rm Ni}}^2}-
\mu_{\rm Ni}ck_{\bot}}{\sqrt{c^2k_{\bot}^2+\mu_{\rm Ni}\omega_{p,{\rm Ni}}^2}+
\mu_{\rm Ni}ck_{\bot}}
\equiv R_{0,{\rm Ni}}(k_{\bot}).
\label{eq16}
\end{equation}
\noindent
Similar to the case of the Drude model [see  Eq.~(\ref{eq8})], the TE reflection
coefficient at zero frequency does not depend on the material of the dielectric
plate, but, unlike the Drude model, it depends on $k_{\bot}$. Substituting
 Eqs.~(\ref{eq15})  and (\ref{eq16}) to the zero-frequency term of Eq.~(\ref{eq1}),
 we have
 \begin{equation}
{\cal F}_{p}^{(l=0)}(a,T)=\frac{k_BT}{4\pi}\int_{0}^{\infty}k_{\bot}dk_{\bot}
\left\{\ln\left[1-e^{-2ak_0^{(2)}(k_{\bot})}\right]+
\ln\left[1-R_{0,{\rm Ni}}^2(k_{\bot})
e^{-2ak_0^{(2)}(k_{\bot})}\right]\right\},
\label{eq17}
\end{equation}
\noindent
where, in accordance to Eq.~(\ref{eq3}),
\begin{equation}
k_0^{(2)}(k_{\bot})=\sqrt{k_{\bot}^2+\mu_{\rm Ni}\frac{\omega_{p,{\rm Ni}}^2}{c^2}}.
\label{eq18}
\end{equation}

As in the case of a nonmagnetic metal film \cite{47}, it can be shown that the quantity
(\ref{eq17}) exponentially fast goes to zero with increasing film thickness. The same is
true for the contribution of all nonzero Matsubara frequencies to the free energy
${\cal F}_{p}^{(l\geq 1)}(a,T)$, where the magnetic permeability should be replaced with
unity, so that the case of magnetic metals is identical to the case of nonmagnetic ones
considered in Ref.~\cite{47}. Note that the zero-frequency contribution (\ref{eq17}) goes
to zero with increasing $a$ even faster, than in the nonmagnetic case, due to an increased
power of exponents [see the multiple $\mu_{\rm Ni}$ in Eq.~(\ref{eq18})].
As a result, the Casimir free energy of a magnetic metal film calculated using the plasma
model approach vanishes exponentially fast with increasing $a$ and does not possess the
classical limit. In the limiting case $\omega_p\to\infty$
the Casimir free energy of a metal film calculated using the plasma
model exponentially fast drops to zero. Thus, the plasma model approach satisfies the
physical requirement that electromagnetic fluctuations cannot penetrate in an interior of
ideal metal.

It is convenient to rearrange  Eq.~(\ref{eq17}) by introducing the new integration variable
$v=2ak_0^{(2)}(k_{\bot})$. In terms of this variable one obtains
\begin{equation}
{\cal F}_{p}^{(l=0)}(a,T)=\frac{k_BT}{16\pi a^2}
\int_{\sqrt{\mu_{\rm Ni}}\tilde{\omega}_{p,{\rm Ni}}}^{\infty}vdv
\left[\ln(1-e^{-v})+\ln(1-R_{0,{\rm Ni}}^2(v)e^{-v})\right],
\label{eq19}
\end{equation}
\noindent
where $\tilde{\omega}_{p,{\rm Ni}}\equiv 2a{\omega}_{p,{\rm Ni}}/c$ and the reflection
coefficient takes the form
\begin{equation}
R_{0,{\rm Ni}}(v)=
\frac{v-\mu_{\rm Ni}\sqrt{v^2-\mu_{\rm Ni}\tilde{\omega}_{p,{\rm Ni}}^2}}{v+
\mu_{\rm Ni}\sqrt{v^2-\mu_{\rm Ni}\tilde{\omega}_{p,{\rm Ni}}^2}}.
\label{eq19a}
\end{equation}

Similar results can be simply obtained for the zero-frequency contribution to the Casimir
pressure of a magnetic film calculated using the plasma model.
Substituting  Eqs.~(\ref{eq15}) and (\ref{eq16}) to the term with $l=0$ in Eq.~(\ref{eq7}),
we have
\begin{eqnarray}
&&
{P}_{p}^{(l=0)}(a,T)=-\frac{k_BT}{2\pi}\int_{0}^{\infty}k_{\bot}k_0^{(2)}(k_{\bot})dk_{\bot}
\label{eq20}\\
&&
\times\left\{
\left[e^{2ak_0^{(2)}(k_{\bot})}-1\right]^{-1}+
\left[R_{0,{\rm Ni}}^{-2}(k_{\bot})e^{2ak_0^{(2)}(k_{\bot})}-1\right]^{-1}\right\}.
\nonumber
\end{eqnarray}
\noindent
In terms of the new variable this equation can be rewritten as
\begin{eqnarray}
&&
{P}_{p}^{(l=0)}(a,T)=-\frac{k_BT}{16\pi a^3}
\int_{\sqrt{\mu_{\rm Ni}}\tilde{\omega}_{p,{\rm Ni}}}^{\infty}v^2dv
\label{eq21}\\
&&
~\times\left\{(e^{v}-1)^{-1}+\left[R_{0,{\rm Ni}}^{-2}(v)e^{v}-1\right]^{-1}\right\}.
\nonumber
\end{eqnarray}
\noindent
All Matsubara terms of the Casimir pressure calculated using the plasma model go to zero
exponentially fast with increasing film thickness \cite{47}. For magnetic metal films
the zero-frequency contribution to the Casimir pressure drops to zero even faster than
for nonmagnetic ones. As a result, the Casimir pressure has no classical limit if the
plasma model is used in calculations.

Now we perform numerical computations of the Casimir free energy and pressure of a Ni film
sandwiched between two sapphire plates or in vacuum using both the Drude and plasma model
approaches. The used dielectric permittivities of Ni and sapphire
along the imaginary frequency
axis are presented in Sec.~II.
The above expressions for the Casimir free energy and pressure of a Ni film, where all
Matsubara terms were expressed in terms of dimensionless variables, have been used.
All computations are performed at room temperature $T=300\,$K.

\begin{figure}[b]
\vspace*{-1cm}
\centerline{\hspace*{0cm}
\includegraphics{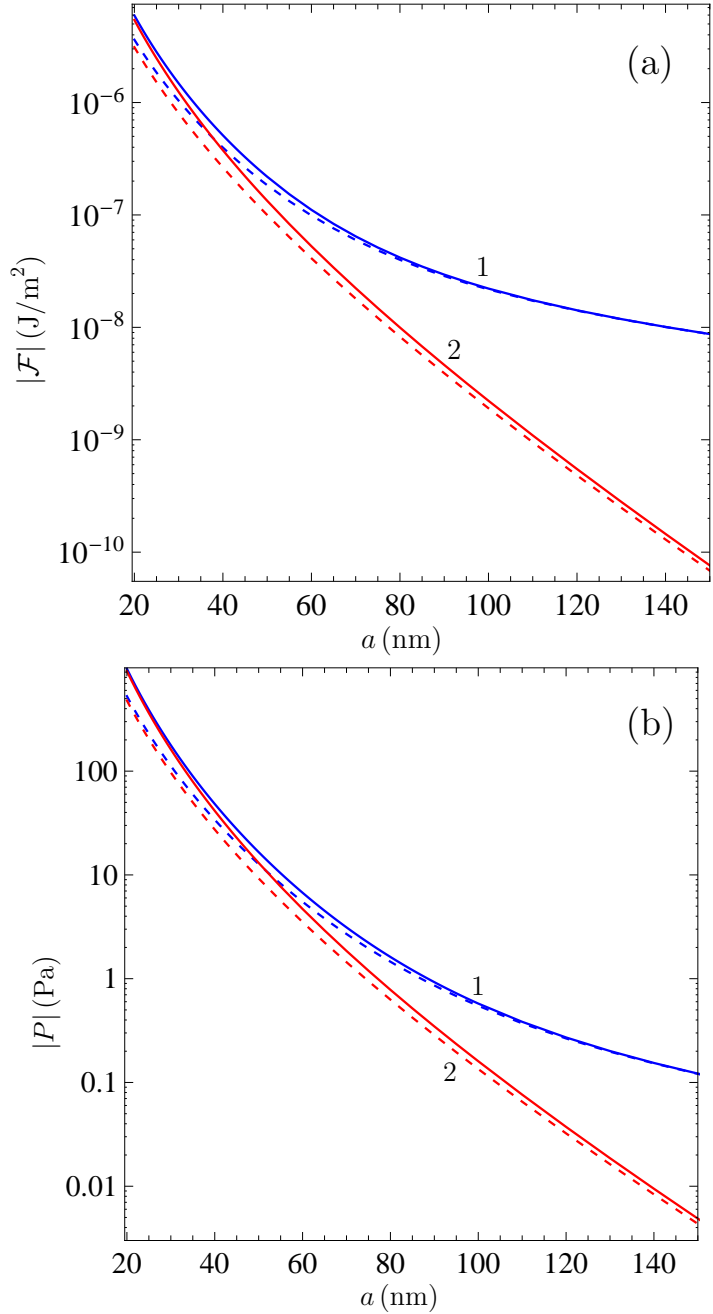}
}
\vspace*{-11cm}
\caption{\label{fg2}(Color online)
The magnitudes of (a) the Casimir free energy per unit area and (b) the Casimir pressure
are computed at $T=300\,$K using the Drude model (solid and dashed lines labeled 1)
and the plasma  model (solid and dashed lines labeled 2) approaches versus the
film thickness for a free-standing Ni film in vacuum (solid lines 1 and 2) and
for a Ni film sandwiched between two sapphire plates (dashed lines 1 and 2).
}
\end{figure}

In Fig.~\ref{fg2}(a) the computational results for the magnitude of the (negative) Casimir
free energy of Ni film in vacuum are shown by the pair of solid lines 1 and 2 as the
functions of film thickness computed using the Drude and plasma model approaches,
respectively. The dashed lines 1 and 2 present similar results for the configuration of Ni
film sandwiched between two sapphire plates. As is seen in Fig.~\ref{fg2}(a), in both
configurations the free energies computed using the Drude model approach (the solid and
dashed lines labeled 1) go to the common classical limit (\ref{eq11}) which is already
reached for a film of 150\,nm thickness.
This corresponds to the characteristic frequency $c/a\approx 1.3\,$eV which is much
larger than the relaxation parameter for Ni (see Sec.~II). One can conclude that the
classical limit is reached in the frequency region of infrared optics where the skin
depth of a Drude model is equal to $\delta=c/\omega_{p,{\rm Ni}}\approx 40\,$nm.
If the plasma model approach is used (the solid and
dashed lines labeled 2) the Casimir free energy in both configurations drops to zero
exponentially fast.

{}From Fig.~\ref{fg2}(a) it is seen that for Ni films of several tens of nanometers
thickness the Drude model approach predicts considerably larger magnitudes of the Casimir
free energy than the plasma model one. Thus, for the Ni films of 50 and 100\,nm thickness
in vacuum (see the solid lines 1 and 2) this excess is by the factors of 1.63 and 9.95,
respectively. Recall that for the configuration of two metallic plates separated with a
vacuum gap of width $a$ the theoretical predictions of both approaches at $a<1\,\mu$m
differ by only a few percent. By the factor of two difference is achieved only at
$a\approx 6\,\mu$m. For the Ni films of 50 and 100\,nm thickness sandwiched between two
sapphire plates  (see the dashed lines 1 and 2) the magnitudes of the Casimir free energy
predicted by the Drude and plasma model approaches differ by the factors of 1.84 and 11.47,
respectively. Note that for the configuration of a Ni film deposited on a sapphire plate
(the second sapphire plate is replaced with a vacuum) the Casimir free energy of a film
is sandwiched between the free energies computed for the two configurations considered
above [see  Fig.~\ref{fg2}(a)].

In Fig.~\ref{fg2}(b) we present
the computational results for the magnitudes of the (negative) Casimir
pressure of Ni film  as the functions of film thickness. As in Fig.~\ref{fg2}(a),
the solid lines labeled 1 and 2 are computed using the Drude and plasma model approaches,
respectively, for the configuration of a free-standing Ni film in vacuum.
The dashed lines 1 and 2 present similar results for a Ni
film sandwiched between two sapphire plates.
As is seen in Fig.~\ref{fg2}(b), the Casimir pressure of the magnetic film possesses the
same characteristic properties, as the Casimir free energy. It is of most importance that
the theoretical predictions of both theoretical approaches differ significantly for
already thin magnetic films and that the Casimir pressures computed using the Drude model
approach (the solid and dashed lines 1) go to the classical limit (\ref{eq14}),
whereas the same pressure computed using the plasma model approach drops to zero
exponentially fast (see the solid and dashed lines 2).

\begin{figure}[b]
\vspace*{-8cm}
\centerline{\hspace*{2cm}
\includegraphics{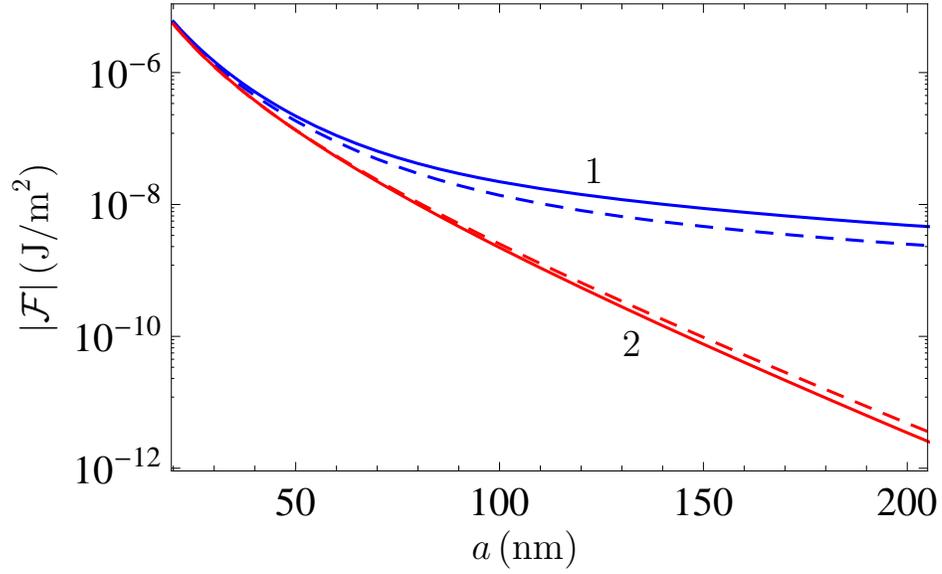}
}
\vspace*{-10cm}
\caption{\label{fg3}(Color online)
The magnitudes of the Casimir free energy per unit area of the
free-standing in vacuum Ni  (solid lines) and Pt (dashed lines) films are
computed at $T=300\,$K  versus the film thickness
using the Drude model (solid and dashed lines labeled 1)
and the plasma  model (solid and dashed lines labeled 2) approaches.
}
\end{figure}

It is interesting to compare the Casimir free energy of a Ni film and a film made of some
nonmagnetic metal, e.g., of Pt, which plasma frequency is approximately equal to that of Ni.
The dielectric permittivity of Pt along the imaginary frequency axis is given
in Sec.~II.
In Fig.~\ref{fg3} the computational results for the magnitudes of the (negative) Casimir
free energy of the free-standing in vacuum Ni (the two solid lines) and Pt (the two dashed lines)
films are shown as the functions of film thickness. The pairs of solid and dashed lines
labeled 1 and 2 are computed using the Drude and plasma model approaches, respectively.
As is seen in Fig.~\ref{fg3}, there is a difference by a factor of two in the classical
limits for the free energies of Ni and Pt films which are already reached at $a=200\,$nm.
This is because the classical limit for the solid line 1 (Ni) is given by Eq.~(\ref{eq11}),
whereas the classical limit for the dashed line 1 (Pt) is given by \cite{47}
\begin{equation}
{\cal F}_{\rm Pt}^{(l=0)}(a,T)=-\frac{k_BT}{16\pi a^2}\zeta(3).
\label{eq22}
\end{equation}
\noindent
When the plasma model approach is used in computations (the dashed and solid lines labeled 2),
the Casimir free energy of both nonmagnetic and magnetic films exponentially fast drops to
zero with increasing film thickness.

\section{Magnetic film deposited on nonmagnetic metal}

In this section we consider a Ni film of thickness $a$ deposited on thick metallic plates made of
either Cu or Al. The role of the third body is now played by a vacuum. In mathematical expressions
below we use the index Cu as an indication of the plate material, but it can be replaced with Al
or any other metal. We start with the contributions of the zero-frequency terms to the Casimir
free energy and pressure calculated using the Drude model approach. Then from
Eq.~(\ref{eq2}) one obtains
\begin{eqnarray}
&&
r_{{\rm TM},D}^{(2,3)}(0,k_{\bot})=-1,
\nonumber \\
&&
r_{{\rm TM},D}^{(2,1)}(0,k_{\bot})=\frac{\omega_{p,{\rm Cu}}^2\gamma_{\rm Ni}-
\omega_{p,{\rm Ni}}^2\gamma_{\rm Cu}}{\omega_{p,{\rm Cu}}^2\gamma_{\rm Ni}+
\omega_{p,{\rm Ni}}^2\gamma_{\rm Cu}}
\equiv R_{0,{\rm NiCu}},
\nonumber \\
&&
r_{{\rm TE},D}^{(2,n)}(0,k_{\bot})=
r_{0,{\rm Ni}},
\label{eq23}
\end{eqnarray}
\noindent
where $r_{0,{\rm Ni}}$ is defined in Eq.~(\ref{eq8}).

Substituting Eq.~(\ref{eq23}) in Eq.~(\ref{eq1}) and using the dimensionless variable
$y=2ak_{\bot}$, we find
\begin{eqnarray}
&&
{\cal F}_{D}^{(l=0)}(a,T)=\frac{k_BT}{16\pi a^2}\int_{0}^{\infty}ydy
\label{eq24}\\
&&
~\times\left[\ln(1+R_{0,{\rm NiCu}}e^{-y})+\ln(1-r_{0,{\rm Ni}}^2e^{-y})\right].
\nonumber
\end{eqnarray}
\noindent
After the integration this results in
\begin{equation}
{\cal F}_{D}^{(l=0)}(a,T)=-\frac{k_BT}{16\pi a^2}\left[{\rm Li}_3(-R_{0,{\rm NiCu}})+
{\rm Li}_3(r_{0,{\rm Ni}}^2)\right].
\label{eq25}
\end{equation}

In a similar way, for the Casimir pressure (\ref{eq7}) one arrives at
\begin{eqnarray}
&&
{P}_{D}^{(l=0)}(a,T)=-\frac{k_BT}{16\pi a^3}\int_{0}^{\infty}y^2dy
\label{eq26}\\
&&
~\times\left[-(R_{0,{\rm NiCu}}^{-1}e^{y}+1)^{-1}+
(r_{0,{\rm Ni}}^{-2}e^{y}-1)^{-1}\right]
\nonumber \\
&&~
=-\frac{k_BT}{8\pi a^3}\left[{\rm Li}_3(-R_{0,{\rm NiCu}})+
{\rm Li}_3(r_{0,{\rm Ni}}^2)\right].
\nonumber
\end{eqnarray}
\noindent
Taking into account the parameters of our configuration, we find from Eqs.~(\ref{eq8}) and
(\ref{eq23}) that $r_{0,{\rm Ni}}=-0.9820$ and $R_{0,{\rm NiCu}}=0.6116$. This results in
\begin{equation}
{\rm Li}_3(-R_{0,{\rm NiCu}})=-0.5716,\quad
{\rm Li}_3(r_{0,{\rm Ni}}^2)=1.1454,
\label{eq27}
\end{equation}
\noindent
i.e., the zero-frequency terms (\ref{eq25}) and (\ref{eq26}) are negative [note that respective
contributions to Eqs.~(\ref{eq11}) and (\ref{eq14}) have similar signs]. As is seen from the
results of numerical computations below, ${\cal F}_{D}^{(l\geq 1)}(a,T)>0$ and for small film
thicknesses it holds
\begin{equation}
|{\cal F}_{D}^{(l\geq 1)}(a,T)|>|{\cal F}_{D}^{(l=0)}(a,T)|.
\label{eq28}
\end{equation}
\noindent
Because of this, the free energy of a Ni film deposited on a nonmagnetic metal plate changes its
sign with increasing film thickness if the Drude model approach is used in
computations
(see below for the values of $a$, where the Casimir free energy
and pressure change sign, for specific plate metals).
Note that this change of sign is not of the same nature, as for
some liquid-separated plates (see, e.g., Refs.~\cite{2,64a}).
In our case the effect is caused by the influence of
magnetic properties on the zero-frequency term when the Drude
model is used in calculations.

Now we calculate the zero-frequency contribution to the free energy and pressure of a Ni film
deposited on a Cu plate using the plasma model. From Eq.~(\ref{eq2}) for the TM reflection
coefficients we find
\begin{eqnarray}
&&
r_{{\rm TM},p}^{(2,3)}(0,k_{\bot})=-1,
\label{eq29} \\
&&
r_{{\rm TM},p}^{(2,1)}(0,k_{\bot})=
\frac{\omega_{p,{\rm Cu}}^2\sqrt{c^2k_{\bot}^2+\mu_{\rm Ni}\omega_{p,{\rm Ni}}^2}-
\omega_{p,{\rm Ni}}^2\sqrt{c^2k_{\bot}^2+\omega_{p,{\rm Cu}}^2}}{\omega_{p,{\rm Cu}}^2
\sqrt{c^2k_{\bot}^2+\mu_{\rm Ni}\omega_{p,{\rm Ni}}^2}+
\omega_{p,{\rm Ni}}^2\sqrt{c^2k_{\bot}^2+\omega_{p,{\rm Cu}}^2}}.
\nonumber
\end{eqnarray}
\noindent
The TE reflection coefficient
$r_{{\rm TE},p}^{(2,3)}(0,k_{\bot})=R_{0,{\rm Ni}}(k_{\bot})$ coincides with that given
by Eq.~(\ref{eq16}) with $n=3$. The second TE reflection coefficient is given by
\begin{equation}
r_{{\rm TE},p}^{(2,1)}(0,k_{\bot})=
\frac{\sqrt{c^2k_{\bot}^2+\mu_{\rm Ni}\omega_{p,{\rm Ni}}^2}-
\mu_{\rm Ni}\sqrt{c^2k_{\bot}^2+\omega_{p,{\rm Cu}}^2}}{\sqrt{c^2k_{\bot}^2+
\mu_{\rm Ni}\omega_{p,{\rm Ni}}^2}+
\mu_{\rm Ni}\sqrt{c^2k_{\bot}^2+\omega_{p,{\rm Cu}}^2}}.
\label{eq30}
\end{equation}

In terms of the variable $v$ the coefficient $r_{{\rm TE},p}^{(2,3)}$ is given by
Eq.~(\ref{eq19a}) and two nontrivial coefficients in Eqs.~(\ref{eq29}) and (\ref{eq30})
take the form
\begin{eqnarray}
&&
r_{{\rm TM},p}^{(2,1)}(0,v)=
\frac{\tilde{\omega}_{p,{\rm Cu}}^2v-\tilde{\omega}_{p,{\rm Ni}}^2
\sqrt{v^2-\mu_{\rm Ni}\tilde{\omega}_{p,{\rm Ni}}^2+
\tilde{\omega}_{p,{\rm Cu}}^2}}{\tilde{\omega}_{p,{\rm Cu}}^2v+\tilde{\omega}_{p,{\rm Ni}}^2
\sqrt{v^2-\mu_{\rm Ni}\tilde{\omega}_{p,{\rm Ni}}^2+
\tilde{\omega}_{p,{\rm Cu}}^2}}
\equiv R_{0,{\rm TM}}(v),
\nonumber  \\[-2mm]
&&\label{eq31}\\[-2mm]
&&
r_{{\rm TE},p}^{(2,1)}(0,v)=
\frac{v-\mu_{\rm Ni}\sqrt{v^2-\mu_{\rm Ni}\tilde{\omega}_{p,{\rm Ni}}^2+
\tilde{\omega}_{p,{\rm Cu}}^2}}{v+\mu_{\rm Ni}
\sqrt{v^2-\mu_{\rm Ni}\tilde{\omega}_{p,{\rm Ni}}^2+
\tilde{\omega}_{p,{\rm Cu}}^2}}
\equiv R_{0,{\rm TE}}(v).
\nonumber
\end{eqnarray}

As a result, the zero-frequency contribution to the Casimir free energy can be written as
\begin{eqnarray}
&&
{\cal F}_{p}^{(l=0)}(a,T)=\frac{k_BT}{16\pi a^2}
\int_{\sqrt{\mu_{\rm Ni}}\tilde{\omega}_{p,{\rm Ni}}}^{\infty}vdv
\label{eq32}\\
&&
\times\left[\ln(1+R_{0,{\rm TM}}(v)e^{-v})+\ln(1-R_{0,{\rm Ni}}(v)R_{0,{\rm TE}}(v)
e^{-v})\right].
\nonumber
\end{eqnarray}
\noindent
The zero-frequency contribution to the Casimir pressure is the following:
\begin{eqnarray}
&&
{P}_{p}^{(l=0)}(a,T)=-\frac{k_BT}{16\pi a^3}
\int_{\sqrt{\mu_{\rm Ni}}\tilde{\omega}_{p,{\rm Ni}}}^{\infty}v^2dv
\label{eq33}\\
&&
\times\left\{-\left[R_{0,{\rm TM}}^{-1}(v)e^{v}+1\right]^{-1}+
\left[R_{0,{\rm Ni}}^{-1}(v)R_{0,{\rm TE}}^{-1}(v)e^{v}-1\right]^{-1}\right\}.
\nonumber
\end{eqnarray}

Numerical computations and the analysis performed in Ref.~\cite{47} show that in the framework
of the plasma model approach
\begin{equation}
|{\cal F}_{p}^{(l=0)}(a,T)|\ll|{\cal F}_{p}^{(l\geq 1)}(a,T)|
\label{eq34}
\end{equation}
\noindent
for any film thickness. Similar to the case of the Drude model, it holds
${\cal F}_{p}^{(l\geq 1)}(a,T)>0$. However, unlike the Drude model, here the sign of
${\cal F}_{p}^{(l\geq 1)}$ determines the positive sign (i.e., the repulsive character)
of the total Casimir free energy of a magnetic film.

\begin{figure}[b]
\vspace*{-2cm}
\centerline{\hspace*{1.cm}
\includegraphics{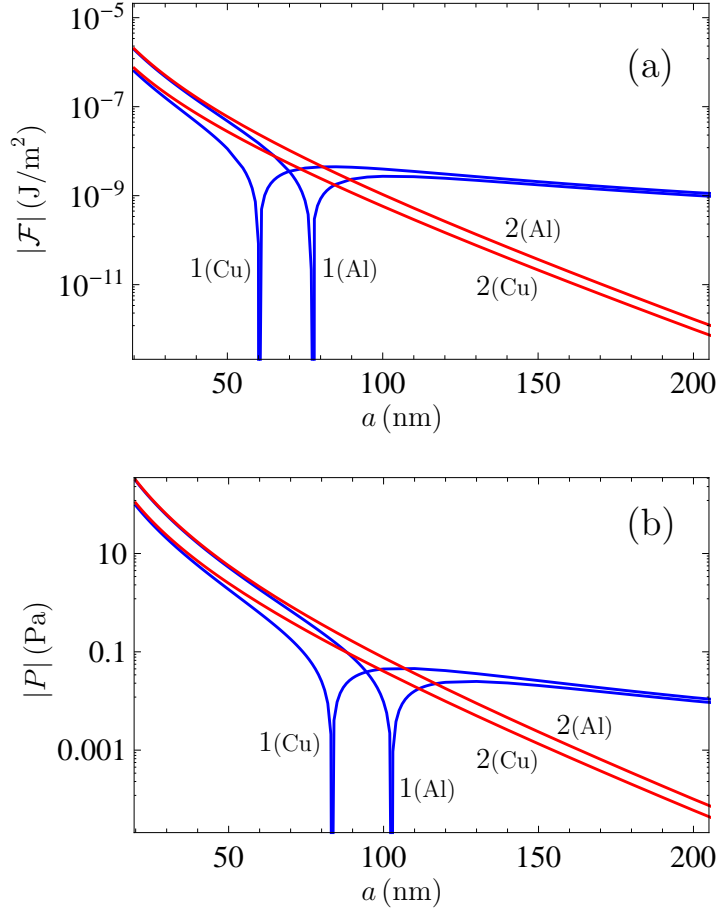}
}
\vspace*{-13cm}
\caption{\label{fg4}(Color online)
The magnitudes of (a) the Casimir free energy per unit area and (b) the Casimir pressure
are computed at $T=300\,$K using the Drude model (lines labeled 1)
and the plasma  model (lines labeled 2) approaches versus the
film thickness for a  Ni film deposited on Cu and Al plates.
If the plasma model is used, the Casimir interaction is repulsive.
In the case of the Drude model, repulsion at short separations
changes for attraction at larger separations.
}
\end{figure}

In Fig.~\ref{fg4}(a) we present the computational results for the magnitudes of the Casimir
free energy of Ni films deposited on Cu and Al plates computed as functions of film thickness
using the Drude model (lines labeled 1) and plasma model (lines labeled 2) approaches.
In Fig.~\ref{fg4}(b) similar results are presented for the Casimir pressure of Ni films.
As is seen from Fig.~\ref{fg4}(a,b), the Casimir free energy and pressure of a Ni film calculated
using the Drude model approach vanish at some film thickness and then change their sign.
At film thickness of approximately 150\,nm both quantities reach their classical limits (\ref{eq25})
and (\ref{eq26}). For the case of Cu plate the Casimir free energy and pressure  are positive for
films thinner than 61.1 and 83.3\,nm, respectively. For the films of these thicknesses
the Casimir free energy and pressure  vanish and become negative for thicker films.
In the case of an Al plate, the Casimir free energy and pressure  calculated using the Drude
model approach take zero value for
$a=77.9$ and 102.7\,nm, respectively. In the limiting case of $\omega_{p,{\rm Ni}}\to\infty$
(the limit of ideal metal)  we have from Eq.~(\ref{eq23}) $R_{0,{\rm NiCu}}\to -1$ and
Eq.~(\ref{eq25}) transforms to  Eq.~(\ref{eq14}) taking into account that ${\rm Li}_3(1)=\zeta(3)$.
This is a nonzero free energy of an ideal metal film, i.e., a counter intuitive result, as
discussed above.

According to Fig.~\ref{fg4}(a,b), quite different results are obtained when the plasma model
approach is used in computations (lines labeled 2). In this case the Casimir free energy and pressure
of Ni films are positive (i.e., correspond to repulsion) and drop exponentially fast to zero with
increasing film thickness. The classical limit is not reached for films of any thickness.
An important point is that theoretical predictions for the magnitudes of the Casimir free energy
and pressure using the Drude and plasma model approaches are significantly different even for
relatively thin Ni films. Thus, for Ni films of 50, 60, and 100\,nm thickness deposited on a Cu
plate the magnitudes of the Casimir free energies computed using both approaches differ by the factors
of 2.44, 140.66 and 6.88, respectively. For Ni films of 50, 75, and 100\,nm thickness deposited on an Al
plate the respective factors are 1.26, 8.98, and 2.56. It is seen that the major difference in
theoretical predictions of the Drude and plasma model approaches occurs for film thicknesses where
the Drude model prediction is close to zero. Similar results are obtained for the Casimir pressure.

\begin{figure}[b]
\vspace*{-8cm}
\centerline{\hspace*{1cm}
\includegraphics{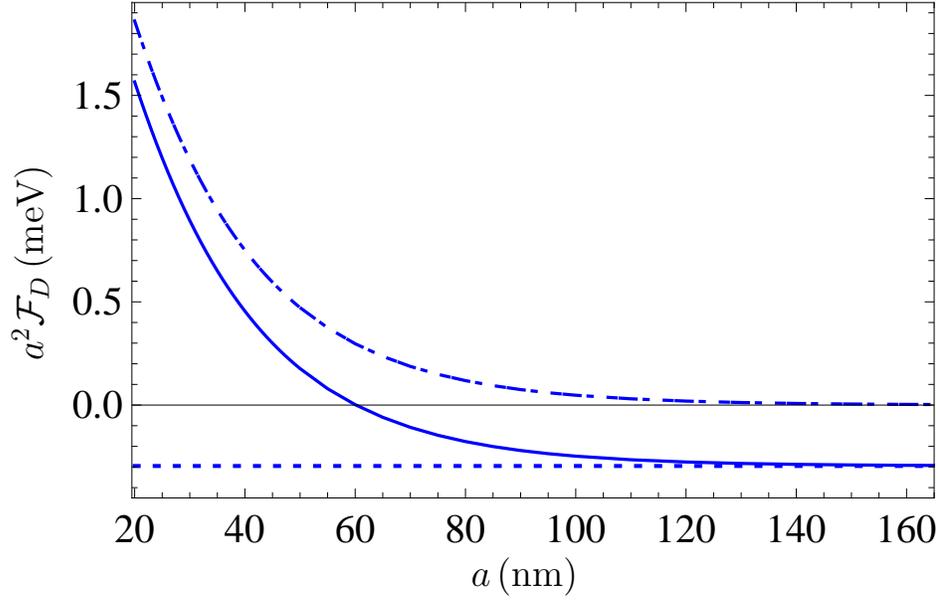}
}
\vspace*{-10cm}
\caption{\label{fg5}(Color online)
The Casimir free energy per unit area of a Ni film deposited on a Cu plate is
computed at $T=300\,$K using the Drude model approach (solid line) versus the
film thickness. The dotted and dashed-dotted lines show contributions to the
free energy of the zero-frequency term and all terms with nonzero
Matsubara frequencies, respectively.
}
\end{figure}

Now we investigate in more detail why the Casimir free energy and pressure change sign as functions
of film thickness if the Drude model approach is used in computations. For this purpose, the solid
line in Fig.~\ref{fg5} shows the quantity $a^2{\cal F}_{D}$ for a Ni film deposited on a Cu plate
as a function of film thickness $a$. It is seen that the free energy changes its sign from positive
to negative in accordance with Fig.~\ref{fg4}(a). In the same figure the dotted line shows the
zero-frequency contribution (\ref{eq32}) to the Casimir free energy, which is a negative quantity.
The dashed-dotted line in Fig.~\ref{fg5} presents the contribution of all
Matsubara terms with nonzero frequencies
to $a^2{\cal F}_{D}$, which is positive for the pair of metals Ni and Cu under
consideration. As a result, the free energy ${\cal F}_{D}$ of a Ni film is positive for
sufficiently thin films and becomes negative with increasing film thickness.

\begin{figure}[b]
\vspace*{-8cm}
\centerline{\hspace*{1cm}
\includegraphics{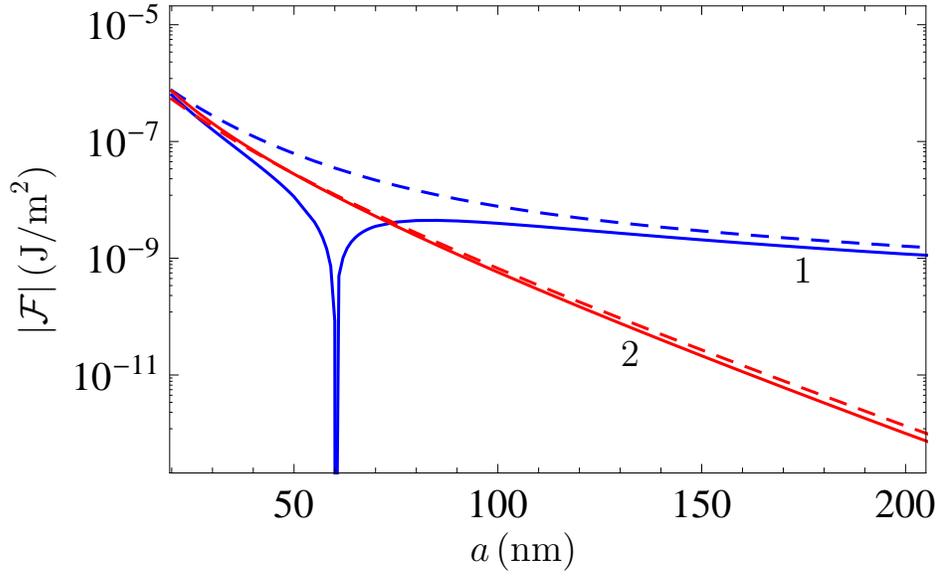}
}
\vspace*{-10cm}
\caption{\label{fg6}(Color online)
The magnitudes of the Casimir free energy per unit area are
computed at $T=300\,$K
using the Drude model (solid and dashed lines labeled 1)
and the plasma  model (solid and dashed lines labeled 2)
approaches versus the film thickness
for a  Ni film  (solid lines 1 and 2) and
for a Pt film (dashed lines 1 and 2) deposited on Cu plates.
The Casimir interaction of Pt film is repulsive for both
the plasma and Drude models.
}
\end{figure}

Next, we consider the role of magnetic properties of the film in the effects discussed above.
For this purpose, in Fig.~\ref{fg6} we reproduce from  Fig.~\ref{fg4} the magnitudes of the Casimir
free energy of a Ni film deposited on a Cu plate computed using the Drude and plasma model
approaches (solid lines 1 and 2, respectively). In the same figure the dashed lines 1 and 2 show the
Casimir free energy of a nonmagnetic Pt film deposited on a Cu plate and computed using the two
approaches. These computations were performed using the dielectric permittivity of Pt presented
on an inset to Fig.~\ref{fg1}(a). The zero-frequency reflection coefficients are given by
Eqs.~(\ref{eq19a}) and (\ref{eq31}) where $\mu_{\rm Ni}$ and $\omega_{p,{\rm Ni}}$ are replaced with
unity and $\omega_{p,{\rm Pt}}$, respectively. As is seen in Fig.~\ref{fg6}, the solid and dashed
lines labeled 2 are similar in appearance, i.e.,  the magnetic properties of a film influence
the Casimir free energy computed using the plasma model approach only quantitatively (note that
for both films ${\cal F}_P>0$). If, however, the Drude model approach is used, this leads to
fundamental differences between the Casimir free energies of nonmagnetic and magnetic films. Thus, the
solid line 1 demonstrates the Casimir free energy of a magnetic Ni film which changes its sign
from positive to negative and takes the zero value for the film of some definite thickness
(see above). At the same time, the Casimir free energy of a nonmagnetic Pt film (the dashed line 1)
is positive for any film thickness. This is also seen from the classical limit for a Pt film
deposited on a Cu plate which is positive
\begin{equation}
{\cal F}_{\rm Pt}^{(l=0)}(a,T)=-\frac{k_BT}{16\pi a^2}{\rm Li}_3(-R_{0,{\rm PtCu}})>0,
\label{eq35}
\end{equation}
\noindent
where $R_{0,{\rm PtCu}}$ is defined by Eq.~(\ref{eq23}) with the index Ni replaced for Pt.
As a result, the use of magnetic films presents additional possibilities for discrimination
between the Drude and plasma model approaches in the Lifshitz theory.

\section{Magnetic film deposited on magnetic metal}

In this section we calculate the Casimir free energy and pressure of a magnetic Ni film deposited
on a plate made of magnetic metal. Numerical computations are performed for a Fe plate using both
the Drude and plasma model approaches. We begin with the Drude model approach. In this case the
reflection coefficients $r_{{\rm TM},D}^{(2,n)}$ at zero Matsubara frequency are again given by
the first two lines of Eq.~(\ref{eq23}) where the index Cu is replaced with Fe.
The reflection coefficient $r_{{\rm TE},D}^{(2,3)}$ at zero frequency is given by the third line
of Eq.~(\ref{eq23}), whereas the coefficient $r_{{\rm TE},D}^{(2,1)}$ is given by
\begin{equation}
r_{{\rm TE},D}^{(2,1)}(0,k_{\bot})=\frac{\mu_{\rm Fe}-\mu_{\rm Ni}}{\mu_{\rm Fe}+\mu_{\rm Ni}}
\equiv r_{0,{\rm NiFe}}>0.
\label{eq36}
\end{equation}

As a result, the zero-frequency contribution to the Casimir free energy of a Ni film on a Fe
plate, calculated using the Drude model, takes the form
\begin{eqnarray}
&&
{\cal F}_{D}^{(l=0)}(a,T)=\frac{k_BT}{16\pi a^2}\int_{0}^{\infty}ydy
\label{eq37}\\
&&
~\times\left[\ln(1+R_{0,{\rm NiFe}}e^{-y})+\ln(1-r_{0,{\rm Ni}}
r_{0,{\rm NiFe}}e^{-y})\right].
\nonumber
\end{eqnarray}
\noindent
Note that $r_{0,{\rm Ni}}r_{0,{\rm NiFe}}<0$. As a result, ${\cal F}_{D}^{(l=0)}>0$,
as opposed to the case of nonmagnetic plate considered in Sec.~IV.
As is seen from the results of numerical computations (see below), the contribution of all
Matsubara terms with nonzero frequency, ${\cal F}_{D}^{(l\geq 1)}$, changes its sign from
plus to minus with increasing film thickness, leaving the total free energy ${\cal F}_D$
positive.
After the integration in Eq.~(\ref{eq37}), one obtains the classical limit
\begin{equation}
{\cal F}_{D}^{(l=0)}(a,T)=-\frac{k_BT}{16\pi a^2}\left[{\rm Li}_3(-R_{0,{\rm NiFe}})+
{\rm Li}_3(r_{0,{\rm Ni}}r_{0,{\rm NiFe}})\right].
\label{eq38}
\end{equation}

In a similar way, for the zero-frequency contribution to the Casimir pressure of a Ni film
on a Fe plate we find
\begin{eqnarray}
&&
{P}_{D}^{(l=0)}(a,T)=-\frac{k_BT}{16\pi a^3}\int_{0}^{\infty}y^2dy
\label{eq39}\\
&&
~\times\left[-(R_{0,{\rm NiFe}}^{-1}e^{y}+1)^{-1}+
(r_{0,{\rm Ni}}^{-1}r_{0,{\rm NiFe}}^{-1}e^{y}-1)^{-1}\right]
\nonumber \\
&&~
=-\frac{k_BT}{8\pi a^3}\left[{\rm Li}_3(-R_{0,{\rm NiFe}})+
{\rm Li}_3(r_{0,{\rm Ni}}r_{0,{\rm NiFe}})\right]>0.
\nonumber
\end{eqnarray}
\noindent

We continue by considering the same configuration using the plasma model approach.
In this case the reflection coefficient $r_{{\rm TM},p}^{(2,3)}$ is given by the first line
of Eq.~(\ref{eq29}) and the coefficient $r_{{\rm TM},p}^{(2,1)}$ is given by
\begin{equation}
r_{{\rm TM},p}^{(2,1)}(0,k_{\bot})=
\frac{\omega_{p,{\rm Fe}}^2\sqrt{c^2k_{\bot}^2+\mu_{\rm Ni}\omega_{p,{\rm Ni}}^2}-
\omega_{p,{\rm Ni}}^2\sqrt{c^2k_{\bot}^2+\mu_{\rm Fe}\omega_{p,{\rm Fe}}^2}}{\omega_{p,{\rm Fe}}^2
\sqrt{c^2k_{\bot}^2+\mu_{\rm Ni}\omega_{p,{\rm Ni}}^2}+
\omega_{p,{\rm Ni}}^2\sqrt{c^2k_{\bot}^2+\mu_{\rm Fe}\omega_{p,{\rm Fe}}^2}}.
\label{eq40}
\end{equation}
\noindent
As to TE reflection coefficients at zero Matsubara frequency,
$r_{{\rm TE},p}^{(2,3)}$ is given by Eq.~(\ref{eq16}) with $n=3$ and
$r_{{\rm TE},p}^{(2,1)}$ takes the form
\begin{equation}
r_{{\rm TE},p}^{(2,1)}(0,k_{\bot})=
\frac{\mu_{\rm Fe}\sqrt{c^2k_{\bot}^2+\mu_{\rm Ni}\omega_{p,{\rm Ni}}^2}-
\mu_{\rm Ni}\sqrt{c^2k_{\bot}^2+\mu_{\rm Fe}\omega_{p,{\rm Fe}}^2}}{\mu_{\rm Fe}
\sqrt{c^2k_{\bot}^2+\mu_{\rm Ni}\omega_{p,{\rm Ni}}^2}+
\mu_{\rm Ni}\sqrt{c^2k_{\bot}^2+\mu_{\rm Fe}\omega_{p,{\rm Fe}}^2}}.
\label{eq41}
\end{equation}
\noindent
In terms of the dimensionless variable $v$ the coefficient $r_{{\rm TE},p}^{(2,3)}$ is given by
Eq.~(\ref{eq19a}) and the coefficients (\ref{eq40}) and (\ref{eq41})
are as follows:
\begin{eqnarray}
&&
r_{{\rm TM},p}^{(2,1)}(0,v)=
\frac{\tilde{\omega}_{p,{\rm Fe}}^2v-\tilde{\omega}_{p,{\rm Ni}}^2
\sqrt{v^2-\mu_{\rm Ni}\tilde{\omega}_{p,{\rm Ni}}^2+\mu_{\rm Fe}
\tilde{\omega}_{p,{\rm Fe}}^2}}{\tilde{\omega}_{p,{\rm Fe}}^2v+\tilde{\omega}_{p,{\rm Ni}}^2
\sqrt{v^2-\mu_{\rm Ni}\tilde{\omega}_{p,{\rm Ni}}^2+
\mu_{\rm Fe}\tilde{\omega}_{p,{\rm Fe}}^2}}
\equiv \tilde{R}_{0,{\rm TM}}(v),
\nonumber  \\[-2mm]
&&\label{eq42}\\[-2mm]
&&
r_{{\rm TE},p}^{(2,1)}(0,v)=
\frac{\mu_{\rm Fe}v-\mu_{\rm Ni}\sqrt{v^2-\mu_{\rm Ni}\tilde{\omega}_{p,{\rm Ni}}^2+
\mu_{\rm Fe}\tilde{\omega}_{p,{\rm Fe}}^2}}{\mu_{\rm Fe}v+\mu_{\rm Ni}
\sqrt{v^2-\mu_{\rm Ni}\tilde{\omega}_{p,{\rm Ni}}^2+
\mu_{\rm Fe}\tilde{\omega}_{p,{\rm Fe}}^2}}
\equiv \tilde{R}_{0,{\rm TE}}(v).
\nonumber
\end{eqnarray}

Then, the zero-frequency contribution to the Casimir free energy and pressure calculated
using the plasma model approach are
\begin{eqnarray}
&&
{\cal F}_{p}^{(l=0)}(a,T)=\frac{k_BT}{16\pi a^2}
\int_{\sqrt{\mu_{\rm Ni}}\tilde{\omega}_{p,{\rm Ni}}}^{\infty}vdv
\nonumber\\
&&
\times\left[\ln(1+\tilde{R}_{0,{\rm TM}}(v)e^{-v})+\ln(1-R_{0,{\rm Ni}}(v)
\tilde{R}_{0,{\rm TE}}(v)
e^{-v})\right],
\nonumber \\
&&
{P}_{p}^{(l=0)}(a,T)=-\frac{k_BT}{16\pi a^3}
\int_{\sqrt{\mu_{\rm Ni}}\tilde{\omega}_{p,{\rm Ni}}}^{\infty}v^2dv
\label{eq43}\\
&&
\times\left\{-\left[\tilde{R}_{0,{\rm TM}}^{-1}(v)e^{v}+1\right]^{-1}+
\left[R_{0,{\rm Ni}}^{-1}(v)\tilde{R}_{0,{\rm TE}}^{-1}(v)e^{v}-1\right]^{-1}\right\}.
\nonumber
\end{eqnarray}
According to the results of Ref.~\cite{47} obtained for nonmagnetic metals,
\begin{equation}
|{\cal F}_{p}^{(l=0)}(a,T)|\ll|{\cal F}_{p}^{(l\geq 1)}(a,T)|
\label{eq44}
\end{equation}
\noindent
for any film thickness $a$. Taking into account that the magnetic properties  only decrease
the magnitude of the zero-frequency terms, Eq.~(\ref{eq44}) is also valid for our case of magnetic metals.

\begin{figure}[b]
\vspace*{-2cm}
\centerline{\hspace*{1.cm}
\includegraphics{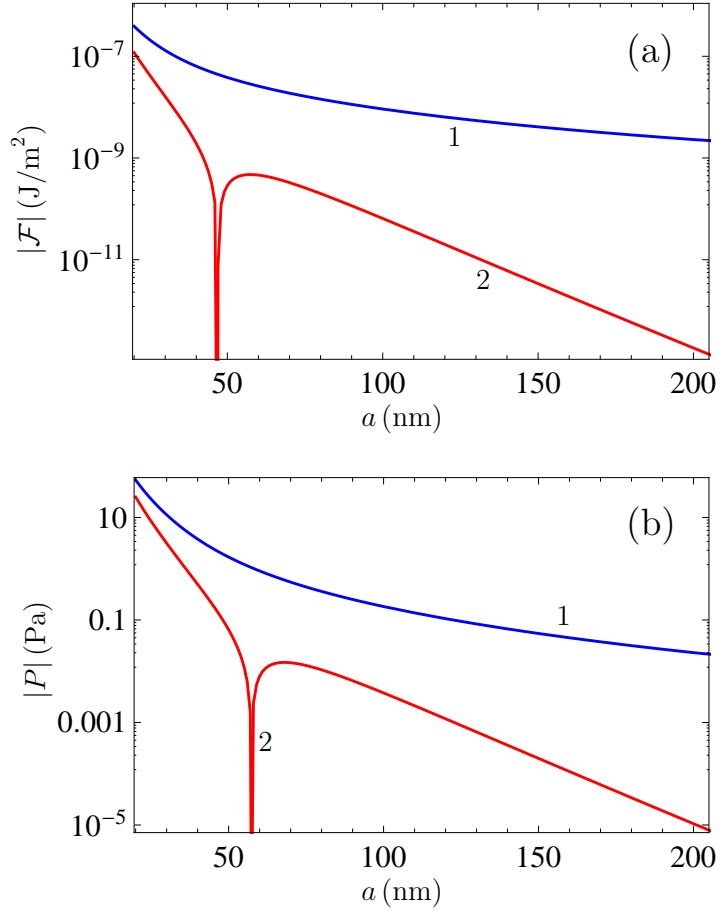}
}
\vspace*{-13cm}
\caption{\label{fg7}(Color online)
The magnitudes of (a) the Casimir free energy per unit area and (b) the Casimir pressure
are computed at $T=300\,$K using the Drude model (lines labeled 1)
and the plasma  model (lines labeled 2) approaches versus the
film thickness for a  Ni film deposited on a Fe plate.
If the plasma model is used, the Casimir interaction is repulsive
at short separations and attractive at larger separations.
In the case of the Drude model, the Casimir interaction is
repulsive.
}
\end{figure}

Numerical computations of the Casimir free energy and pressure for a Ni film deposited on a Fe plate
were performed using the Drude and plasma model approaches with the dielectric permittivity of Fe
along the imaginary frequency axis presented in Fig.~\ref{fg1}(b). In Fig.~\ref{fg7}(a) we present the
computational results for the magnitude of the Casimir free energy as a function of film thickness
computed using the Drude model (line 1) and plasma model (line 2) approaches. As is seen in
Fig.~\ref{fg7}(a), the free energy ${\cal F}_D$ goes to its classical limit (\ref{eq39}), whereas
the free energy ${\cal F}_p$ exponentially fast drops to zero. The free energy ${\cal F}_D$  is
positive for all $a$, and  the free energy ${\cal F}_p$  is positive for sufficiently thin films,
takes zero value for the film of 46.95\,nm thickness and becomes negative for thicker films.
This is just the opposite to Figs.~\ref{fg4}(a) and \ref{fg6}, where the Casimir free energy
computed using the Drude model approach changes its sign, whereas the one calculated using the
plasma model remains positive.

The next important feature of Fig.~\ref{fg7} is that the differences in theoretical predictions of
the Drude and plasma models are much larger than for a plate made of nonmagnetic metal. Thus, for
Ni films of 47, 50, and 100\,nm thickness the magnitudes of the Casimir free energy predicted using
both theoretical approaches are different by the factors of 5866, 133.3, and 142.0, respectively.
The largest difference for a film of 47\,nm thickness is explained by the fact that
for a film of 46.95\,nm thickness ${\cal F}_p=0$.

The computational results for the magnitudes of the Casimir pressure are presented in Fig.~\ref{fg7}(b)
by the line 1 (the Drude model approach) and line 2 (the plasma model approach)  as the functions of
film thickness. The line 1 corresponds to the Casimir repulsion and follows the classical limit
(\ref{eq39}) for Ni films of more than 150\,nm thickness. The line 2 describes the Casimir repulsion
for the films of less than 57.4\,nm thickness, where the Casimir pressure computed using the plasma
model takes zero value. For the films of larger thickness the Casimir pressure $P_p$ takes negative
values and drops to zero exponentially fast. In Fig.~\ref{fg7}(b) the differences in the Casimir
pressures predicted by the Drude and plasma model approaches are much larger than those in
Figs.~\ref{fg2}(b) and \ref{fg4}(b). Because of this, the magnetic plate is advantageous for making
discrimination between the two approaches.

\begin{figure}[b]
\vspace*{-8cm}
\centerline{\hspace*{1cm}
\includegraphics{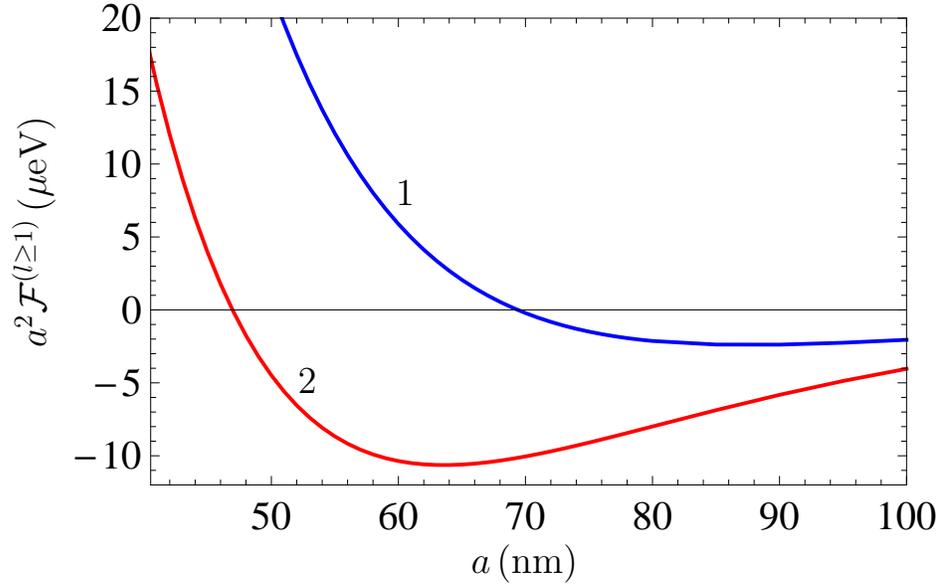}
}
\vspace*{-10cm}
\caption{\label{fg8}(Color online)
The contribution of all nonzero Matsubara frequencies to the
Casimir free energy per unit area of a Ni film deposited on a Fe plate is
computed at $T=300\,$K using the Drude model (line 1) and the plasma
model (line 2) approaches versus the film thickness.
}
\end{figure}

Now we illustrate the role of the Matsubara terms with nonzero frequencies in the complete Casimir
free energies in Fig.~\ref{fg7}(a) computed using the Drude and plasma model approaches.
In Fig.~\ref{fg8} we present the computational results for the quantity $a^2{\cal F}^{(l\geq 1)}(a,T)$
computed using the Drude model (line 1) and plasma model (line 2) approaches as the functions of
film thickness. As is seen in Fig.~\ref{fg8}, the contribution of all nonzero Matsubara frequencies
changes its sign for both theoretical approaches.  In the case of the Drude model approach, however,
the contribution of the zero-frequency term (\ref{eq38}) remains much larger in magnitude:
$a^2{\cal F}_{D}^{(l=0)}(a,T)=575.6\,\mu$eV. This explains why the total free energy computed using the
Drude model approach remains positive for Ni films of any film thickness. If the plasma model
approach is used (line 2), the contribution of the zero-frequency Matsubara term is negligibly small
in accordance to Eq.~(\ref{eq44}). This is the reason why the complete Casimir free energy
${\cal F}_p(a,T)$ changes its sign with increasing film thickness.

\section{Conclusions and discussion}

In the foregoing we have calculated the Casimir free energy and pressure for a magnetic metal film
(Ni) sandwiched between two dielectric plates (sapphire) or deposited  on a plate made of either
nonmagnetic (Cu, Al) or magnetic (Fe) metal. All calculations have been performed at room temperature
using the tabulated optical data of all metals under consideration in the framework of both the Drude
and plasma model approaches to the Lifshitz theory of the van der Waals and Casimir forces.
This is important in connection with the problem in Casimir physics
concerning the role  of dissipation of
free charge carriers discussed in Sec.~I.

According to our results, the Casimir free energy and pressure of magnetic films differ widely
depending on the calculation approach used for already rather thin films. For the free-standing, sandwiched
between two sapphire plates or deposited on Cu and Al plates Ni films of several tens nanometer
thickness these differences reach hundreds and thousands percent. For a Ni film of 47\,nm thickness
deposited on a plate made of magnetic metal (Fe) the Casimir free energies predicted using the
Drude and plasma approaches differ by the factor of 5866. This opens opportunities for an
unambiguous experimental discrimination between the two approaches with no resort to the differential
force measurements.

We have shown that the Casimir free energy and pressure of a magnetic film calculated using the
Drude model approach possess the classical limit in all configurations considered above, which is
reached at about 150\,nm film thickness. If the plasma model approach is used in calculations,
the Casimir free energy and pressure have no classical limit, but, instead, exponentially fast drop
to zero with increasing film thickness. The absence of the classical limit, however, is not a
disadvantage of the plasma model approach, but more likely is its merit. The point is that for
a metallic film the presence of the classical limit results in a nonzero Casimir free energy even
though the metal of a film becomes an ideal one. This result should be considered as an indication
of failure because the fluctuating electromagnetic filed cannot penetrate in an interior of ideal
metal. The reason for so strong influence of magnetic properties when the Drude model approach is
used is that they act only through the zero-frequency term of the Lifshitz formula. The latter is
dominant in the Drude model approach and negligibly small in the plasma model one, as compared
to the contribution of all nonzero Matsubara frequencies.

The numerical computations show that if the Drude model approach is used the Casimir free energy
and pressure of a Ni film deposited on nonmagnetic plates made of Cu and Al change their sign with
increasing film thickness. In the framework of the plasma model approach these quantities are
positive for all film thicknesses, i.e., correspond to the Casimir repulsion. The situation changes
if a Ni film is deposited on a magnetic Fe plate. In this case the Casimir free energy and pressure
are positive if the Drude model approach is used and change their sign with
increasing film thickness if calculations are performed using the plasma model approach.
All these results are important not only for the discrimination between existing theoretical
approaches, but also in the investigation of stability of thin films.

The possibility to directly measure the single-film Casimir pressure is discussed in Ref.~\cite{47}.
For this purpose, the most often used sphere-plate geometry can be employed, where the solid
metallic film is replaced with a liquid metal like mercury or, more conveniently, with an alloy of
gallium and indium. In our case the magnetic liquid metals should be used \cite{61}.
The predicted effects can be also observed in the Casimir chips \cite{19}, where it is not
difficult to maintain parallelity. In this case the Casimir pressure in a magnetic film is
measurable by means of an additional piezo sensor.

An answer to the question why the plasma model approach to the Casimir force is consistent with
experimental data and thermodynamics though the plasma model does not describe the optical
properties and conductivity of metals at low frequencies might be rooted in the foundations of
quantum statistical physics. The optical properties and conductivity of metals are measured as
a response of a system to the electromagnetic fields with nonzero expectation values.
The usually used postulate that the dielectric response of metals to the fluctuating fields having
a zero expectation value is precisely the same is some kind of extrapolation, which works good in classical
statistical physics, but might be not applicable to all kinds of low-frequency quantum fluctuations.
Future investigation will show whether this approach is helpful for the resolution of the
problems in Casimir physics.


\newpage
\end{document}